# COVID-19: THE INFORMATION WARFARE PARADIGM SHIFT

Jan Kallberg, PhD[1,2], Rosemary A. Burk, PhD[3], and Bhavani Thuraisingham, PhD[4].

[1] *Army Cyber Institute at West Point, the U.S. Military Academy, West Point, New York*

[2] *Department of Social Sciences, the U.S. Military Academy, West Point, New York*

[3] *U.S. Fish and Wildlife Service, the Department of the Interior, Head Quarters, Falls Church, Virginia*

[4] *Cyber Security Institute, Erik Jonsson School of Engineering and Computer Science, The University of Texas at Dallas, Richardson, Texas*

Corresponding author: Jan Kallberg jan.kallberg@westpoint.edu

# COVID-19: THE INFORMATION WARFARE PARADIGM SHIFT

**Introduction**

In Kuhn's "The Structure of Scientific Revolutions,"[1] the critical term is paradigm-shift when it suddenly becomes evident that earlier assumptions no longer are correct – and the plurality of the scientific community that studies this domain accepts the change. These types of events can be scientific findings or as in social science system shock that creates a punctured equilibrium that sets the stage in the developments.

In information warfare, recent years' studies and government lines of efforts have been to engage fake news, electoral interference, and fight extremist social media as the primary combat theater in the information space, and the tools to influence a targeted audience. The COVID-19 pandemic generates a rebuttal of these assumptions. Even if fake news and extremist social media content may exploit fault lines in our society and create a civil disturbance, tensions between federal and local government, and massive protests, it is still effects that impact a part of the population. What we have seen with COVID-19, as an indicator, is that what is related to public health is far more powerful to swing public sentiment and create reactions within the citizenry that are trigger impact at a larger magnitude that has rippled through society in multiple directions. These ripple effects have been hard to predict. The long-term psychological, societal, and health impacts of these events have still not yet unfolded. As an example, according to the National Bureau of Economic Research, no other historic pandemic event has affected the stock market as profoundly as COVID-19.[2]

COVID-19 has provided an essential data set to compare what matters to the population. The environmental aspect of cyber defense, linked to public health, has not drawn attention as a national security matter. We all, as living beings, react to threats to our living space and near environment. Jeopardizing the environment, unintended or intended, has historically led to the immediate injection of fear and strong reactions in the population. Even unanticipated accidents with environmental impact have triggered strong moves in the public sentiment towards fear, panic, anger against the government, and challenges to public authority.

In retrospect, we can always formulate excellent explanations, but we can also test these assumptions logically. From the advisory's perspective, what impact can they have on a presidential election, and does it matter if a Democratic or Republican President elected? What is the upside? The U.S. defense spending and grand outlook on the world order have been almost consistent over the decades. Even if presidents and political leaders have made broad statements of swift moves in different policy directions, the actual change in the geopolitical landscape has been marginal. As a recent example, President Trump's movement of troops from Germany to Poland, Belgium, and Italy is instead a rearrangement and a geopolitical new position. From a Russian perspective, with an increasingly more military able Poland and increasing commitment from several NATO countries, the U.S. move of troops out of Germany does not change the current situation. The return on the Russian information warfare investment is not present.

According to Waltz, it is not what you do, but instead what you can do that gives you the power.[3] An adversary can gain more influence over the popular sentiment through threatening to harm the immediate environment and public health, and all major potential adversaries to the U.S. do not subscribe to the ethics, code of conduct, and playbook as we do. COVID-19 has shown that cyber attacks that create environmental and health threats, even with very low probability to occur, creates drastic swings in the sentiment. Cyber attacks that threaten public health and the citizens' immediate environment put the targeted government's legitimacy, authority, and control under pressure and trigger a significant decrease in the citizens' confidence in the current political leadership. The magnitude of such impact can hardly be created by tweets, fake news, and rally extremists on social media because these events can be proven false and are perishable in the public eye, but plausible threats to health and environment last.

Humans have survived through thousands of years by learning, remembering, and adapting to avoid threats to life and health. Therefore, cyber-attacks that trigger fears of threats to public health and personal life has not only a massive initial impact but also lasting effects that migrate to general perception and policy.

One such example is the Three Mile Island accident that created significant public turbulence and fear – an incident that still has a profound impact on how we envision nuclear power. For a covert state actor that seek to cripple our society, embarrass the political leadership, and project to the world that we cannot defend ourselves, environmental damages are inviting.[4] An attack on the environment feels for the general public more close and scary than a dozen servers malfunctioning in a server park. It is tangible and quickly becomes personable and relatable, beyond what politically incendiary memes and social media storms can create.

We are all dependent on clean drinking water and non-toxic air. Cyber attacks on these fundamentals for life could create panic and desperation in the general public – even if the reacting citizens were not directly affected.[5]

The last decade's study of cyber has left the environmental risk posed by cyber-controlled networks unaddressed.[6] The focus on cybersecurity has included providing for restoration of information systems by incorporating detection, protective, and reactive capabilities. From the information security's early inception in the 1980s to today's secured environments, we have become skilled in our ability to secure and harden information systems. The interest in critical infrastructure is to a high degree of accessibility, dependency, and availability that the systems are working and restoring their working condition after an attack. Instead, the long-lasting impact of the threat to human health and the immediate environment drives sentiment and impact policy further by a concerned citizenry than a temporary loss of service. The environmental effects would be dramatic and long-term; freshwater resources contaminated, complete ecosystems destroyed, toxic agents released, and massive soil erosion. Environmental damages and threats to our immediate environment are tangible and highly visible - flooding, undrinkable water, pandemic, biological hazards, mudslides, toxic air, and chemical spills directly affect the population and their surrounding environment. A failed computer server park does not drive media attention, nor can a few hundred tweets create such an impact on the public sentiment as a hundred thousand dead fishes floating down a river. The environmental impact is visible, connects with people on a visceral level, and generates a notion that the human core existence is in jeopardy. Humans put survival first.

Environmental damages trigger radical shifts in the public mind and the general sentiment. For a minor state actor, such as an adversarial developing nation, these attacks can be done with limited budget and resources and still create significant political turbulence and loss of confidence in the population of a targeted major actor. Conflict and potential war, as mentioned, seeks to change policy and influence another nation to take steps that it earlier was unwilling to do. The panic that can follow environmental damages is a political force worth recognizing, which COVID-19 has evidenced. Systematic cyber-attacks that threatens public health will likely generate influence with enough momentum to change national policy.

*Loss of Legitimacy and Authority*

Covert successful cyberattacks that lead to environmental impact are troublesome for the government – not only the damage – but also the challenge to legitimacy, authority, and confidence in the government and political leadership. The citizens expect the state to protect them. The protection of the citizenry is one of the core elements in the concept of a democratic government. The security of the citizens is a part of the unwritten social contract between then citizens and the government. The federal government's ability to protect is taken for granted – it is assumed to be in place. If the government fails to protect and safeguard the citizens, the legitimacy is challenged—legitimacy concerns not who can lead but who can govern. A failure to protect is an inability to govern the nation entrusted, and legitimacy is eroded. Institution stability can be affected and destabilize the nation. The political scientist Dwight Waldo believed that we need faith in government; for the government to have a strong legitimacy, it has to project, deliver, and promise that life would be better for citizens. In a democracy, the voter needs a sense that they are represented, government works for their best, and government improves life for citizens and voters. In the "Administrative State," Waldo defined his vision of the "good life" as the best possible life for the population that can be achieved based on the time, technology, and resources.[7] Authority is the ability to implement policy.

Environmental hazards that lead to loss of life and dramatic long-term loss of life quality for citizens trigger a demand for the government to act. If the population questions the government's ability to protect and safeguard the government's legitimacy and authority will suffer. In the Three Mile Island accident, the event had an impact on sentiment and risk perception, even decades after the incident, on how citizen's perceived the government's nuclear policies and ability to ensure that nuclear power was safe.

President Carter needed to show and project the ability to handle the incident and to restore confidence in the general public for the government's policies. Environmental risks tend to appeal not only to our general public's logic but also emotions, foremost to the notion of uncertainty and fear, and a population that fears the future has instantly lost confidence in the government.

The difference with the Three Mile Island incident and cyber attacks on our infrastructure, creating environmental damage, is that the Three Mile Island incident was local, solitaire could be contained and understood. During the Three Mile Island incidents, millions of Americans had a real fear for their life and future –when faced with the possibility of a nuclear meltdown.

Cyber attacks on our national infrastructure in pursuit of threats to public health cannot be predicted or contained. These attacks can be massive if the exploit utilized for the attack target a shared vulnerability. So the fear generated by Three Mile Island could, in retrospect, have been marginal to the fear caused by a large scale cyber attack on the national infrastructure.

*Environmental cyber defense*

Defending American infrastructure from cyberattacks is not only protecting information, network availability, or the global information grid. It is also safeguarding the public health and the environment, which affects the lives of citizens, their health, their immediate living environment, and protecting ecosystems that we rely upon. Attacks on the immediate environment and the quality of life of the citizenry directly affect the confidence the population has in the government's ability to govern at a magnitude that was visualized by the COVID-19 epidemic.

---

[1] Kuhn, Thomas S. The structure of scientific revolutions. University of Chicago press, 2012.

[2] Baker, Scott R., Nicholas Bloom, Steven J. Davis, Kyle J. Kost, Marco C. Sammon, and Tasaneeya Viratyosin. The unprecedented stock market impact of COVID-19. No. w26945. National Bureau of Economic Research, 2020.

[3] Waltz, Kenneth N., 'Nuclear Myths and Political Realities', American Political Science Review, (Sept. 1990), 731-745.

[4] Kallberg, Jan, and Rosemary A. Burk. "Failed Cyberdefense: The Environmental Consequences of Hostile Acts." Military Review 94, no. 3 (2014): 22.

[5] Kallberg, Jan., and Rosemary A. Burk (2013). Cyber Defense as Environmental Protection—The Broader Potential Impact of Failed Defensive Counter Cyber Operations in *Conflict and Cooperation in Cyberspace-The Challenge to National Security in Cyberspace*. Eds. A., Yannakogeorgos, P. A., & Lowther, A. B. (2013). Taylor & Francis: New York.

[6] Idaho National Laboratory. 2005. US-CERT Control Systems Security Center. Cyber Incidents Involving Control Systems. INL/EXT-05-00671.
http://www.inl.gov/technicalpublications/documents/3480144.pdf

[7] Waldo, Dwight. 1980. *The enterprise of public administration*. Novato: Chandler & Sharp.